# Towards Resolving Landauer's Paradox Through Direct Observation of Multiscale Ferroelastic-Ferroelectric Interplay

*Colm Durkan,[1] Asaf Hershkovitz[2,3] DaPing Chu,[4] James F. Scott[5] and Yachin Ivry[2,3]\**

*Correspondence to: ivry@technion.ac.il

1. Nanoscience Centre, University of Cambridge, 11 JJ Thomson Ave., Cambridge, CB3 0FF, UK.

2. Department of Materials Science & Engineering, Technion – Israel Institute of Technology, Haifa, 32000, Israel.

3. Solid State Institute, Technion – Israel Institute of Technology, Haifa, 32000, Israel.

4. Electrical Engineering Division, University of Cambridge, 9 JJ Thomson Ave., Cambridge, CB3 0FA, UK.

5. School of Chemistry, University of St Andrews, Purdie Building, North Haugh, St Andrews, KY16 9ST, UK.

ABSTRACT

Electric-polarization reversibility in nano-ferroelectric structures renders them as a convenient platform for exploring phase transitions and developing energy-efficient switching devices. However, the fundamental question of how ferroic domains switch, i.e. how the polarization changes from one state to another, is yet to be answered fully. There are contradicting models and a wide body of accumulated data which disagree as to whether the switching requires domain nucleation. Moreover, ferroelectric domains switch under electric fields that are supposedly too weak to form nucleation sites, indicating that the level of disorder seen in real systems plays an important role. This longstanding so-called *Landauer's paradox* is the ferroelectric equivalent to the absence of raindrop formation in a dust-free vacuum, leading to supersaturated vapors that cannot exist otherwise, e.g. in spinodal decompositions or inhomogeneous nucleation environments. Here we show that polarization switching in ferroelectric-ferroelastic systems comprises domain types that differ by symmetry, lengthscale and switching energy. These domains switch simultaneously thanks to intermediate-range order of organized pinning sites, supporting the previously-unexplained coexistence of nucleation-and-growth and nucleation-frustrated mechanisms. Our treatment is applicable to other Kolmogorov-Avrami systems with multi-scale phase transitions. Finally, we demonstrate augmented electromechanical coupling based on the collective motion of pinning sites, which is promising for nano electro-mechanical and low-power switching devices.

There is extensive recent literature on self-patterning and faceting and the resulting geometry of domain walls, especially for thin-film ferroelectrics and other two-dimensional systems. This high level of interest is driven by the fact that, as Berge *et al*. have emphasized [1], faceting is not permitted at thermal equilibrium in two dimensions. Because the perimeter of a two-dimensional structure is one-dimensional, thermodynamically it cannot exhibit long-range order at finite temperatures [2,3]. However, two-dimensional faceting can occur dynamically during domain switching processes and has been modelled numerically for growth previously [4]. In the present work we take a fresh look at domain geometries in ferroelectric thin films that are repetitively switched. We build on earlier work by Du and Chen [5] (expanded by Tagantsev [6] and Hesse, and Alexe [7]), Kalinin [8], Gruverman [9], Waser, [10] and Salje [11], emphasizing comparisons with Kolmogorov-Avrami-Ishibashi (KAI) [12] switching (controlled by domain wall mobility) and nucleation-limited switching (NLS) models [5,6] in non-uniaxial ferroelectric crystals. Our results show that ferroelectric domain switching is a multiscale process with a strong dependence on the ferroelastic-ferroelectric interplay.

We focus in particular on biferroic systems, where the characteristic length scales for polarization correlation (ferroelectricity) and strain (ferroelasticity) differ significantly, being typically of the order 1 micron and 10 nm, respectively. This leads to a hierarchy of mesoscopic structures, some of which exhibit pseudo symmetry, *i.e.* domain symmetries higher than those of the underlying crystallographic lattice. Such mesoscopic block arrays (e.g. polytwins) can qualitatively change the switching dynamics from the atomic-scale creep of domain walls in low-defect systems (observed first by Little [13]) to array-jumping, reminiscent of Barkhausen jumps in magnetism. Tybell *et al.* showed [14] how domain-wall creep exponents depend upon defects in ferroelectrics, but did not consider jumps of entire blocks. A convenient framework to discuss the array jump is "bundle" domains [15], where the ferroelastic polytwins help define a mesoscopic-scale ferroelectric behavior. However, other polytwin systems typical to ferroelectric-ferroelastic systems also exhibit similar collective behavior. An analogous bundle-jumping is the well-known 1962 Anderson model of Abrikosov vortex motion in Type-II superconductors [16,17]. In each case, an unknown (but repulsive) intra-site force is involved. In addition, bearing in mind the highly-anisotropic one-dimensional nature of ferroic domain walls, low-mobility ferroic domain switching may follow de Gennes' "repetition" model in polymers, in which one-dimensional chains slide like spaghetti through each other. Similar low-wall mobility has recently been attributed to observations of oxygen vacancy footprints [18], which in turn can lead to unconventional electric and mechanical coupling with magnetism [19].

The essence of ferroelectricity is the reversibility of polarization under an external electric field. Therefore, the most important and fundamental question to answer in ferroics is: *is there a single model to describe the domain switching*? Having such a model is also technologically significant, *e.g.* for commonly-used low-power switching devices [20]. Traditionally, two competing mechanisms have been suggested to explain this switching. The first is the KAI model, in which domain switching is equivalent to a phase transition in an infinite space. In this model, application of an external electric field gives rise to randomly distributed nucleation points. Upon increasing the electric field, these nuclei merge to form a single domain with common macroscopic polarization. Quantitatively, the fraction of the switched area satisfies:

$$p(t) = 1-\exp(-(t/t_0)^n) \qquad (1)$$

where $t_0$ is the switching time and $n$ is the dimension ($n = 2$ for thin films). The KAI model assumes a constant nucleation growth rate. However, experimental results, especially those involving high electric fields, challenge this assumption [6,7]. Hence, an alternative NLS approach was also suggested. Here, the domain switching occurs when the polarization in many uncorrelated elementary regions is reversed independently [5,6] and the fraction of the switched area satisfies:

$$p(t) = 1 - \langle \gamma_i \exp(-(t/\tau_i)) \rangle_i \qquad (2)$$

where $1/\tau_i$ is the nucleation rate within an elementary region $i$, $\gamma_i$ is the area of region $i$ normalised by the average area of elementary regions in the ferroelectric and $\langle ... \rangle_i$ is the average over the different regions $i$.

Gruverman *et al*. observed [9] polarization switching in lead zirconate titanate (PZT) directly, by means of piezoresponse force microscopy (PFM [21–23]). Specifically, they measured the evolution of an area under gradually-switched polarization as well as the number of nucleation sites as a function of switching time. Their data shows that the switching involves two distinguishable steps and they were able to quantitatively fit the first step to the KAI model and the second step to the NLS model. Hence, a decisive conclusion about which of the switching mechanisms governs the polarization reversal has remained elusive, as the two models are incompatible. We offer here an alternative interpretation to these results. We suggest that indeed, there are several independent switching processes. However, these processes take place simultaneously, albeit at different length-scales. Specifically, by introducing a multiscale ferroelectric-ferroelastic interplay, we explain the complex polarization reversal mechanism in ferroelectrics that is consistent with the presumably contradicting existing models and experimental results. The relationship between length-scale and switching time have been proposed for various cases, including for nucleation and growth [24–26]. However, here we suggest a more general relationship between the two parameters. We note that although most ferroelectrics also exhibit ferroelasticity due to a change in the crystal class during the paraelectric-ferroelectric transition, some uniaxial ferroelectrics do not [27], and hence require a more subtle treatment [28].

Most ferroelectrics, especially in thin-film form are composed of different crystallographic domains. Such domains arise to compensate for stress accumulated during the ferroelectric phase transition due to, *e.g.*, substrate clamping. As a result, out-of-plane polarization switching involves both electric and electro-mechanical interactions, and thus, at least two different switching mechanisms. We have recently demonstrated that mesoscopic domains with ordered arrays of crystallographic domains should be treated as individual ferroelectric domain entities [15], as exemplified in Figure 1. Such mesoscopic structures can include polytwins or mash ferroelastic-ferroelectric platform. Here we focus on bundle domains. It has been shown that such bundle domains can switch as a whole, producing a net out-of-plane mesoscopic polarization reversal [26,29,30]. Generally, the bundle-domain size is comparable to or larger than the film thickness [31], but is smaller than the electrodes used to apply the electric field (this is valid for all ferroelectric-based devices and for most relevant experimental studies of domain switching kinetics in thin films). When discussing domain switching, one should take into account both the out-of-plane motion of the atomic dipoles (180° switching) and the reorientation of the ferroelastic twinning planes associated with bundle switching. Although these two switching mechanisms are different in nature [32,33], the switching of mesoscopic/macroscopic out-of-plane polarization involves both of them.

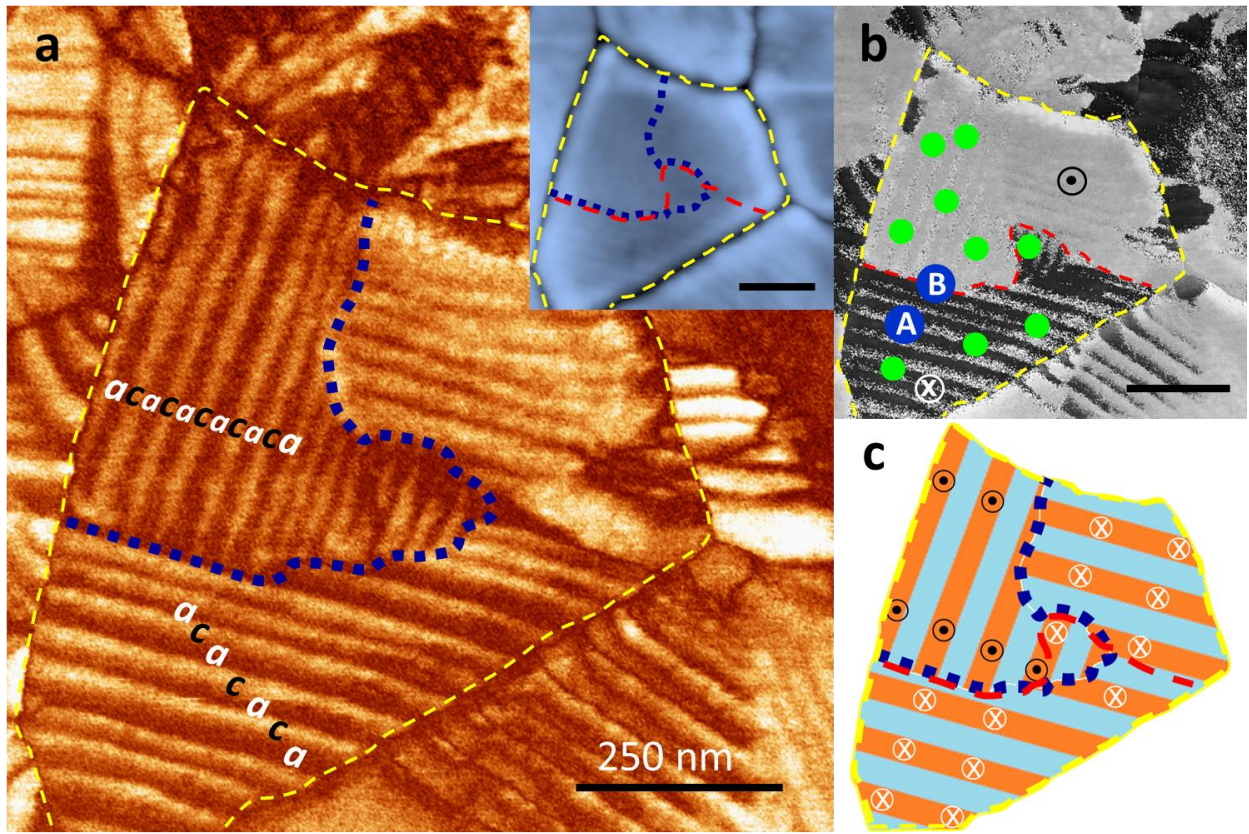

**Figure 1| Correlated pinning sites and mesoscopic polarization distribution**. (**a**) Ferroelastic domains (observed with enhanced-PFM amplitude mode) are arranged in bundle domains, *i.e.* sets of aligned stripes with alternating crystallographic orientation (designated by *a* and *c*); (**b**) The simultaneously-imaged mesoscopic out-of-plane polarization (EPFM phase mode) is arranged in mesoscopic structures. Green and blue dots mark the points where local hysteresis loop measurements were measured; (**c**) Schematics of the domain distributions in the main grain. Here, out-of-plane domains are colored in orange and other crystallographic orientations in blue. Bundle domains are distinguished with a dashed blue line and mesoscopic polarization boundaries are marked with a dashed red line. Inset in (a): the simultaneously-imaged topography [using atomic force microscopy (AFM) imaging] shows the grain distribution, as well as some of the striped ferroelastic domains (the bundle domain boundary is discussed in more detail in Ref. [15]).

The existence of such a dual mechanism that involves both 180° and ferroelastic domain switching is supported *e.g.* by the unexplained two-step switching mechanism observed by Gruverman *et al*. [9] and more recently by Khan *et al*. [33]. The creep description of the domain kinetics is valid for fast and continuous motions [14,34]. In ceramics, creep occurs at ~40-50% of the phase transition temperature. The ferroelectric transition in PZT takes place at ~750 K, suggesting that domain creep occurs already at room temperature and creep dynamics describe the fast and continuous KAI-like domain switching. However, it might be useful to consider the process that occurs before creep, *i.e.* at lower speeds. Prior to creep, there is a very slow depinning regime. As the depinning rate increases, eventually the regime normally considered as creep will take place (in a quasi-continuous transition process). In such a pre-creep regime, each depinning step can be regarded as a nucleation event. Thus, the pre-creep depinning regime can be described by the NLS, in which bundle domains are analogous to the elementary regions of this model, while the true creep regime is described by the KAI model. Finally, a bundle-domain boundary is the meeting point for a large number of correlated defects and pinning centers [35] The association of multiple pinning centers with the NLS model was originally proposed by Tagantsev et al. [6] as a possible alternative explanation for the NLS formalism they used. That is, in the presence of multiple pinning sites, the exponent in Eq. 1 should be averaged for a broad range of $t_0$:

$$p(t) = 1 - \langle \exp(-(t/t_0)) \rangle_{t_0} \tag{3}$$

Given that Eq. 2 can be approximated by Eq. 3 easily, the presence of multiple pinning sites gives rise to the same mathematical expression that describes the NLS. In the creep framework, the rate of change of the switched area with time, *dp/dt* is the relevant parameter rather than *p(t)*). Hence, by taking the time derivative of the data from Gruverman *et al*. [9] an obvious separation between the two processes is distinguished on a *dp/dt* vs. *t* graph (Fig. 2).

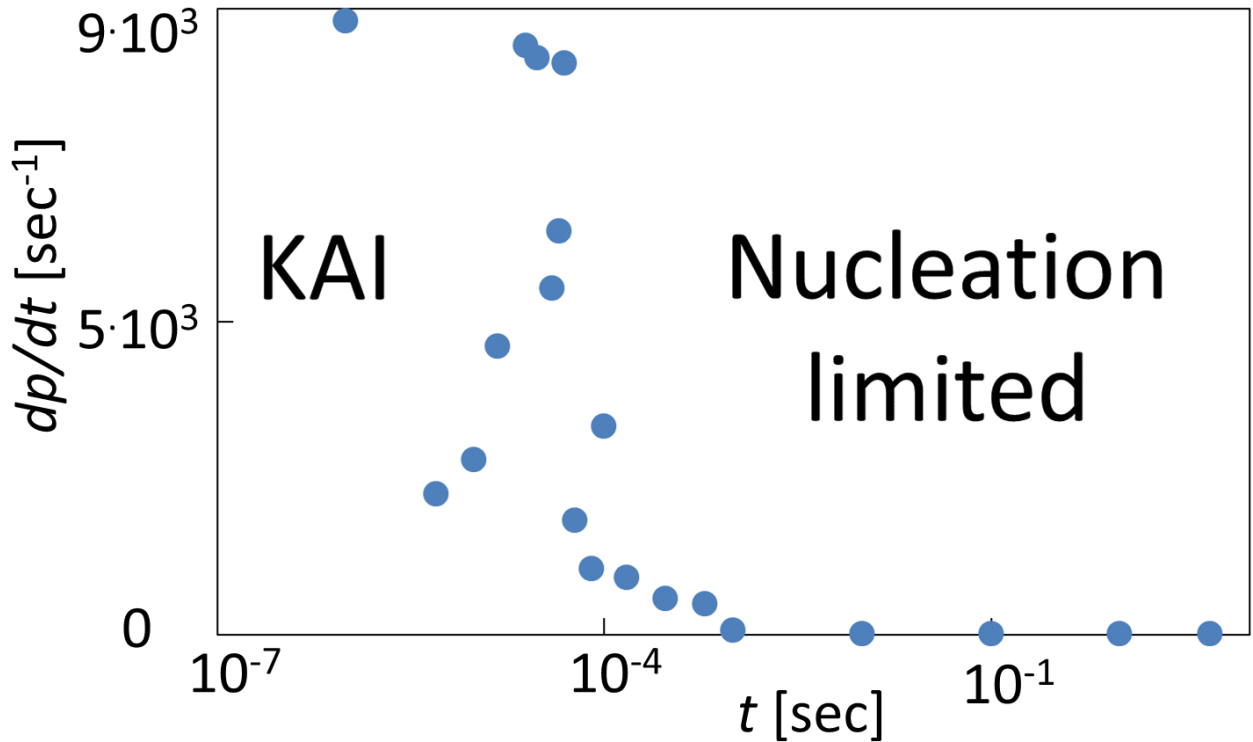

**Figure 2| Multiple processes in domain switching**. Rate of change of the fraction of the switched area ($dp/dt$) as a function of time in PZT demonstrates a clear separation between different switching processes with a sharp decrease at $5\text{-}8\cdot10^{-5}$ sec associated with a distinction between KAI and homogenous (nucleation-frustrated) switching behavior. Data reproduced [36] from Fig. 2b (1.1 V dataset) of Reference [9].

The domain-wall length is generally much larger than the extent of individual pinning sites, so that the average influence of the latter is negligible [6]. Nevertheless, in bundle domains, the pinning sites are ordered (meeting points between three or more crystallographic orientations), and they significantly influence the domain stability and mobility [35]. Hence, we can conclude that the switching process in ferroelectric-ferroelastic systems, such as PZT, takes place simultaneously over two primary length-scales. The first is a local fast 180° domain switching, described by the KAI model and the second is a mesoscopic bundle-domain switching that is accompanied by a broad distribution of ordered elastic-domain wall pinning sites that complies with the NLS (or by the mathematical equivalent of multiple KAI switching processes).

A possible explanation for correlated pinning sites (e.g. polytwins or bundle domains) being switched as individual entities is that the piezoelectric effect acts asymmetrically on neighboring crystallographic domains. For instance, for PZT, $d_{33}$ and $d_{13}$ are 125 and 60 pC/N [37], respectively, allowing 50% strain difference between *a* and *c* domains under a given applied voltage. The corresponding strain may locally exceed the Poisson ratio, requiring stress release which results in a rearrangement of the polytwins, leading to domain switching. Moreover, the dissimilar twinning periodicity within different bundle domains and the variation in bundle domain types [31] can contribute towards the distribution of $t_0$. In terms of creep, the ordered twinning defects at a boundary of a polytwin or a bundle domain, or even in a mash ferroic structure are likely to result in dislocation pinning [35]. Here, there is no explicit dependence on domain size, but a strong dependence on the differential stress across the boundary (although the number of pinning sites is proportional to the bundle-domain perimeter). Such dislocation creep also clarifies the longstanding unexplained observation by Lohse *et al*. [10] of broad relaxation time distributions and of rounded switched domain walls that are typical to Frank-Read sources pinning sites, which in turn have been reported in the context of mesoscopic domain boundaries [38]. Likewise, we would like to suggest that the KAI switching is associated with Coble creep of a symmetrically-growing domain due to diffusion across the domain boundary.. A difficulty with the KAI model is that experimentally, an increase of the applied field does not change the shape of $p(t)$ as predicted by this model, rather, it shifts the curve along the time axis [6,9]. We would like to propose that this shift results from a delay in the switching initiation. Following Harrison and Salje [11], the initiation of domain switching occurs as an avalanche, with a probability depending on the applied field (recent studies suggest a different behavior when non-180⁰ domains are absent [39]). We found this model in agreement with recent experimental studies that ferroelastic and ferroelectric domain switching occur at different voltages [33]. Hence, the shift in switching initiation with applied field represents the change in probability to avalanche, which is also mediated by ferroelastic domain switching.

To verify the two-step domain switching mechanism by means of direct observation, one has to demonstrate that (i) both local 180° and mesoscopic bundle-domain switching can occur simultaneously in the ferroic film; and (ii) the energy associated with the two switching processes is different (following Jesse *et al*. [8]). We examined the influence of ordered pinning sites on the domain motion in PZT films[40] via enhanced piezoresponse force microscopy (EPFM) [41]. In EPFM, the polarization of ferroelectric domains is imaged in phase-mode (areas with in-plane polarization appear with a lower contrast) and the ferroelastic domains (non-180°) are imaged in amplitude-mode (ferroelectric domain walls appear as closed dark lines). The topography is also simultaneously mapped in AFM mode, in which the periodic ferroelastic domains are occasionally distinguishable. Because the domains are imaged at nanoscale resolution [41], and no top electrode short-circuits the domains, domains of different lengthscale are imaged and switched independently. Figure 3 illustrates that the proposed KAI-like step is possible, *i.e.* out-of-plane polarization is switched locally within a bundle domain without moving twinning planes. We first used amplitude and phase EPFM modes for imaging the native striped ferroelastic-domain (Fig. 1a) and out-of-plane polarization distribution (Fig. 1b) in a grain that is much larger than the 60 nm film thickness (and hence is considered as a single crystallite). The crystallite contained two main bundle domains, while the out-of-plane polarization (Fig.1b) divided the grain into two obvious mesoscopic areas of 'up' and 'down' oriented polarization (such bundle-domain boundary type was discussed elsewhere [31]).

With piezoresponse microscopy, one can sweep the voltage ($V_{tip}$) between the tip and a bottom electrode upon which the ferroelectric film is deposited, while recording the ac deflection of the cantilever. A local hysteresis loop measurement (LHLM) simultaneously switches the domain and measures the switching energy, enabling direct observation of the switching energy associated [8]. We first performed LHLMs within a bundle domain, aiming to switch only the out-of-plane component of polarization, without affecting the ferroelastic domain structure. Figures 3a-b demonstrate the domain distribution as a result of the LHLMs within the bundle domains (*i.e.* $V_{tip}$ was applied locally for LHLMs at the points designated by small green dots and point A in Fig. 1b). Subsequent EPFM imaging confirms that only the out-of-plane ferroelectric (180°) domains were switched (Fig. 3b), while the ferroelastic domains (stripes) remained unchanged (Fig. 3a).

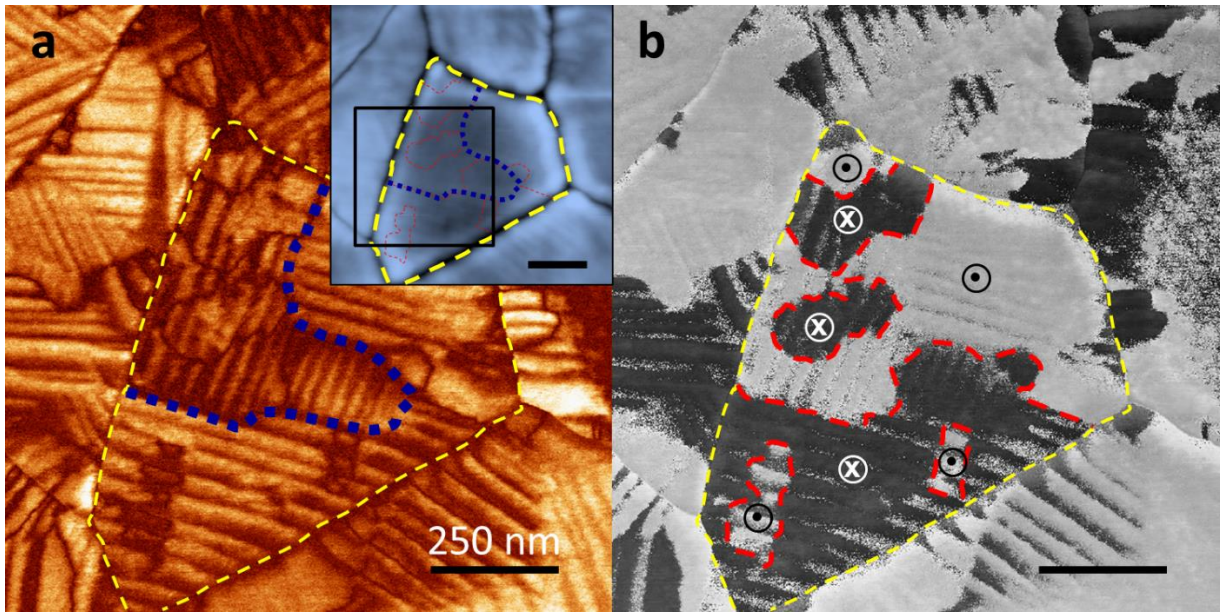

**Figure 3| Effect of local polarization switching on the ferroelastic-ferroelectric interplay. (a)** The ferroelastic domain distribution (imaged with EPFM amplitude mode) in the crystallite of interest was not affected by the local polarization switching at chose points within the bundle domains after LHLMs were measured at the points designated by green dots and the point "A" in Fig. 1b; **(b)** The simultaneously-imaged polarization distribution (EPFM phase signal) demonstrates that the LHLMs switched the out-of-plane component of polarization (the depolarized ferroelectric domain walls appear as closed dark lines in (a)). Bundle-, polarization- and grain- boundaries are highlighted with dashed lines (blue, red and yellow, respectively), while the simultaneously-imaged topography is given in the inset in (a).

When the same excitation is applied to a bundle domain boundary, the reversal in the out-of-plane polarization is accompanied by ferroelastic domain switching, which in turn is associated with a specific switching energy. An external field was then applied while positioning the tip at the bundle-domain boundary, *i.e.* exciting the domains exactly at the pinning sites, where the domain switching is prone to nucleate (point B in Fig. 1b) [26]. Indeed, the external excitation this time moved the ferroelastic domains, *i.e.* the bundle domain wall(Fig. 4).

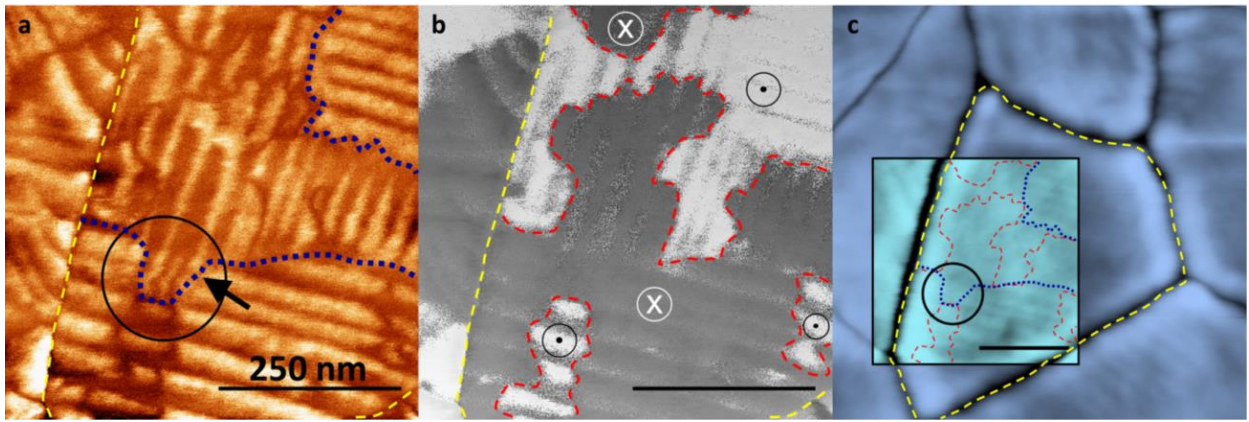

**Figure 4| Local depinning of a ferroelectric-ferroelastic bundle-domain wall.** (**a**) Ferroelastic domain distribution in the region highlighted in Fig. 3a (EPFM amplitude signal) after performing LHLM at the bundle domain boundary (point 'B' in Fig. 1b). Twinned stripes from one bundle domain were depinned, expanding to the neighboring bundle domain (the area of interest is marked with a circle and an arrow). (**b**) The simultaneously-imaged out-of-plane polarization (EPFM phase signal) shows that the 180° polarization domains were also switched locally as a result of the depinning. (**c**) The simultaneous topography mapping of the area (highlighted square) within the large grain shows many of the striped ferroelastic domains. Dashed lines (blue, red and yellow) designate bundle-, polarization- and grain- boundaries, respectively.

LHLMs allow one to compare the energy associated with domain switching. Although there have been attempts to attribute energetic differences to simultaneous motion of ferroelectric and ferroelastic domains, typically, this comparison has not been done in conjunction with the ferroelectric and ferroelastic domain switching imaging [42,43]. However, by comparing the EPFM images of the native domain distribution and the redistribution as a result of the LHLMs, we were able to identify the energy profile with a specific switching process. The electro-mechanical energy (the area within the hysteresis curve) associated with the two domain-switching mechanisms--bundle-domain switching (Fig. 5a), and 180° domain switching only (Fig. 5b)--is compared. The recorded in-phase signal LHLMs are given in Fig. SI1. A clear difference between the energy of these two processes is observed. Our direct characterisation of hysteresis behavior as a result of the different domain motion also supports previous suggestions based on indirect measurements of electromechanical augmentation due to non-180° domain motion by, *e.g.* Nagarajan and co-authors [42,44] and LeRhun *et al.* [43] in agreement with earlier discussions by Arlt [45].

The imaging enhancement of EPFM requires the usage of soft cantilevers (<< 10 N/m) [41] Using such soft cantilevers reduces the confidence in quantitative analysis of the piezoresponse signal in terms of absolute values, but does not affect the confidence in comparative measurements [46,47]. To complete the quantitative analysis, we also performed a macroscopic hysteresis loop measurement, which is complementary to the LHLMs (Fig. 5). Here, we swept the electric field in a plate capacitor that includes the substrate of the sample and a 1 mm diameter top Pt electrode. The plate capacitor was connected to a 198 nf capacitor in a Sawyer-Tower scheme, so that the polarization ($P$) was measured as a function of the applied electric field ($E$) as shown in Fig. 5c.

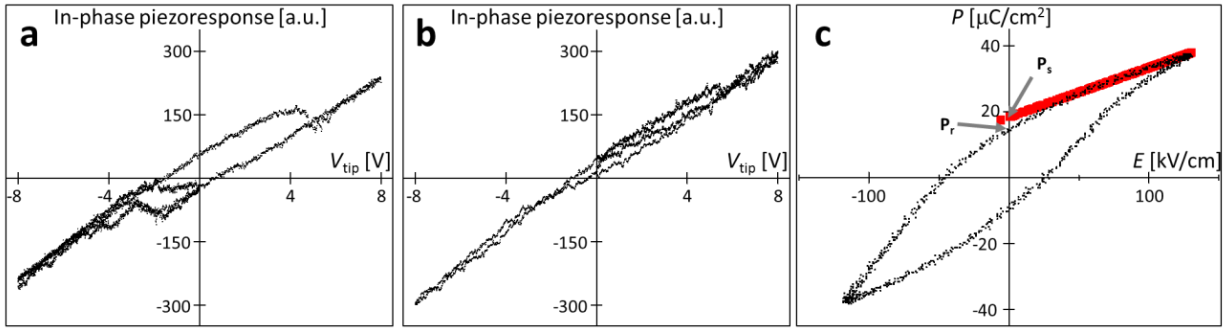

**Figure 5| Multiscale hysteretic behavior.** LHLM of the in-phase piezoresponse signal as a function of applied voltage revealed a larger electro-mechanical coupling (**a**) during bundle domain switching (at point 'B' in Fig. 1b) than (**b**) during 180° switching in a center of a bundle domain (at point 'A' in Fig. 1b). (**c**) Macroscopic polarization-electric field loop measured with a Sawyer Tower scheme. The extrapolated linear line in (c) demonstrates the difference between $P_s$ (= 18.4 ± 0.4 µC/cm$^2$) and $P_r$ (≈ 14.5 µC/cm$^2$). The resultant free energy satisfies: $\Delta G(P) \approx -1.7 \cdot 10^7 P^2 + 2.5 \cdot 10^8 P^4$ ($P$ is in V/m). The higher value of $P_s$ in (a) with respect to (b) suggests that bundle domains are pinned stronger than 180° domains. As well as that, the area of the loop in (a) is more than twice as big as that in (b), suggesting that the electromechanical energy associated

with the switching is around twice as big – perhaps the ability to rotate domains in-plane will be useful for energy harvesting applications. In-phase LHLMs are given in Fig. SI1.

Figure 5c allowed us to extract a quantitative description of the macroscopic free energy of our samples. Given the two possible dipole moment orientations in each unit cell (up and down), the simplest symmetric expression for the free energy is: $\Delta G(P) = 0.5\chi_1 P^2 + 0.25\chi_{11} P^4 - EP$, where the $\chi$'s are different order of dielectric stiffness, $E$ is an external electric field, and we neglected higher-order terms). The requirement for minimum energy dictates: $\partial G/\partial P = 0$ [48], so that the spontaneous polarization ($P_s$)--the value of the polarization in the absence of external electric field--satisfies:

$$P_s^2 = -\chi_1/\chi_{11} \qquad (4)$$

and the dielectric stiffness (the inverse permittivity) can be extracted from the derivative of the electric field at $P_s$: $\partial E(P = P_s)/\partial P = -2\chi_1$. In ideal ferroelectrics, $P_s$ is extrapolated from the linear behavior at high electric field values and is distinguished from the remnant polarization value at $E = 0$ ($P_r$) as denoted in Fig. 5c. Hence, the linear extrapolation in Fig. 5c satisfies:

$P = P_s + (-2\chi_1)^{-1} E$ with $P_s = 18.4 \pm 0.4$ µC/cm² and $\chi_1 = -3.33 \cdot 10^7$. Substituting these values in Eq. 4 indicates that the free energy of our system at the macroscopic scale is given by: $\Delta G(P) \approx -1.7 \cdot 10^7 P^2 + 2.5 \cdot 10^8 P^4$, where $P$ is given in V/m. Although the use of soft cantilevers and the consequent complex factors contributing to the cantilever deflection do not permit a similar straightforward derivation at the meso- and nano- scale $P_s$ is larger for the case of non-180° switching (Fig. 5a). Bearing in mind Eq. 4, this suggests that the potential barrier to be overcome during switching is larger in this case than in 180° domain switching, emphasising the pinning associated with bundle domain walls.

In summary, we demonstrated that true ferroelectric domain switching occurs when both 180° and bundle-domain switching take place simultaneously, but at different length-scales. This gives rise to a two-step switching process that comprises a fast KAI and a slow NLS behavior, while we interpreted the latter as an averaging of multiple KAI kinetic events from ordered depinning sites. Such a multiscale ferroelastic-mediated ferroelectric domain switching process can explain the previous observations of the coexistence of several apparently contradicting switching mechanisms: nucleation-and-growth and nucleation-frustrated. These domain switching mechanisms take place simultaneously, although each at their own timescale. We propose that such a multiscale domain switching mechanism can be used to resolve Landauer's paradox, which concerns polarization reversion and does not consider the motion of correlated pinning sites as those we report here in the case of bundle-domain boundary. Finally, we have demonstrated by means of direct observation piezoelectric and hysteretic augmentation of bundle-domain switching with respect to 180° polarization reversal.

In addition to the analysis discussed here, we note that our suggestion that a switching event occurs simultaneously at two time and length scales that span several orders of magnitude has been reported recently by Stolichnov *et al*. [18] Here, the authors reported that domain walls attract oxygen vacancies, which in turn give rise to electric conducting paths. When applying an external electric field for a short time (a few ns) the switched domain walls move. Yet, the trail or footprint of oxygen vacancies remains for a day or two before diffusing away. This activity of oxygen vacancies extends our analysis beyond ferroelectricity and ferroelasticity, to magnetic domain walls: In a perovskite titanate, the higher likelihood of the presence of oxygen vacancies in domain walls converts neighboring $Ti^{+4}$ to $Ti^{+3}$, while $Ti^{+3}$ is magnetically active. Therefore, there will be enhanced magnetism at the domain walls even in PZT. This magnetism interacts with the ferroelastic strain via magnetostriction of the form of E2H2 (or P2M2), producing a magnetodielectric response [19]. In these scenarios, many of the effects depend strongly on the oxygen vacancies and can therefore be reduced, *e.g.*, by annealing in oxygen, allowing future examination of our analysis. Finally, combined with the recently reported controllable bundle domain reversal [26], the observed electromechanical coupling augmentation is expected to advance ferroic-based nano electro-mechanical systems, such as actuator, cellular antennae etc as well as nanoscale switching devices to lower-power and higher performance.

**Methods**

Imaging and LHLMs were carried out using an Asylum MFP-3D, with Pt/Cr coated ContE cantilevers from Budget-Sensors, resonant frequency: 11–14 kHz, nominal force constant: 0.2 N/m. Details about the EPFM method can be found elsewhere [29,41,47, 49, 50] as well as details about the PZT films [40]. The images were analysed with WsXM [51].

**References**

[1]     B. Berge, L. Faucheux, K. Schwab, A. Libchaber, *Nature* **1991**, *350*, 322.
[2]     L. D. Landau, E. M. Lifshitz, *Statistical Physics, Volume 5*; Elsevier (imprint: Butterworth-Heinemann): Oxfors, 1996.
[3]     I. Lukyanchuk, P. Sharma, T. Nakajima, S. Okamura, J. F. Scott, A. Gruverman, *Nano Lett.* **2014**, *14*, 6931.
[4]     R. Savit, R. Ziff, *Phys. Rev. Lett.* **1985**, *55*, 2515.
[5]     X. Du, I.-W. Chen, *Ferroelectrics* **1998**, *208-209*, 237.
[6]     A. Tagantsev, I. Stolichnov, N. Setter, J. Cross, M. Tsukada, *Phys. Rev. B* **2002**, *66*, 214109.

[7]     Y. Kim, H. Han, W. Lee, S. Baik, D. Hesse, M. Alexe, *Nano Lett.* **2010**, *10*, 1266.

[8]     S. Jesse, B. J. Rodriguez, S. Choudhury, A. P. Baddorf, I. Vrejoiu, D. Hesse, M. Alexe, E. a Eliseev, A. N. Morozovska, J. Zhang, L.-Q. Chen, S. V Kalinin, *Nat. Mater.* **2008**, *7*, 209.

[9]     A. Gruverman, B. J. Rodriguez, C. Dehoff, J. D. Waldrep, a. I. Kingon, R. J. Nemanich, J. S. Cross, *Appl. Phys. Lett.* **2005**, *87*, 082902.

[10]    O. Lohse, M. Grossmann, U. Boettger, D. Bolten, R. Waser, *J. Appl. Phys.* **2001**, *89*, 2332.

[11]    R. J. Harrison, E. K. H. Salje, *Appl. Phys. Lett.* **2010**, *97*, 021907.

[12]    Y. Ishibashi, Y. Takagi, *J. Phys. Soc. Japan* **1971**, *31*, 506.

[13]    E. Little, *Phys. Rev.* **1955**, *98*, 978.

[14]    T. Tybell, P. Paruch, T. Giamarchi, J.-M. Triscone, *Phys. Rev. Lett.* **2002**, *89*, 097601.

[15]    Y. Ivry, D. P. Chu, C. Durkan, *Nanotechnology* **2010**, *21*, 065702.

[16]    P. W. Anderson, *Phys. Rev. Lett.* **1962**, *9*, 309.

[17]    A. M. Troyanovski, J. Aarts, P. H. Kes, **1999**, *399*, 665.

[18]    I. Stolichnov, M. Iwanowska, E. Colla, B. Ziegler, I. Gaponenko, P. Paruch, M. Huijben, G. Rijnders, N. Setter, *Appl. Phys. Lett.* **2014**, *104*, 132902.

[19]    G. Catalan, *Appl. Phys. Lett.* **2006**, *102902*, 88.

[20]    J. F. Scott, *Science* **2007**, *315*, 954.

[21]    A. Gruverman, *J. Vac. Sci. Technol. B Microelectron. Nanom. Struct.* **1995**, *13*, 1095.

[22]    P. Güthner, K. Dransfeld, *Appl. Phys. Lett.* **1992**, *61*, 1137.

[23]    C. Durkan, M. Welland, D. Chu, P. Migliorato, *Phys. Rev. B* **1999**, *60*, 16198.

[24]    H. M. Duiker, P. D. Beale, J. F. Scott, C. A. Paz de Araujo, B. M. Melnick, J. D. Cuchiaro, L. D. McMillan, *J. Appl. Phys.* **1990**, *68*, 5783.

[25]    P. K. Larsen, G. L. M. Kampschöer, M. J. E. Ulenaers, G. A. C. M. Cuppens, R. Spierings, *Appl. Phys. Lett.* **1991**, *59*, 611.

[26]    Y. Ivry, J. F. Scott, E. K. H. Salje, C. Durkan, *Phys. Rev. B* **2012**, *86*, 205428.

[27]    J. C. Toledano, J. Schneck, *Solid State Commun.* **1975**, *16*, 1101.

[28]    V. Y. Shur, In *Nucleation Theory and Applications*; Schmelze, J. W. P., Ed.; WILEY-VCH (Weinheim), 2005; pp. 178–214.

[29]    Y. Ivry, N. Wang, D. Chu, C. Durkan, *Phys. Rev. B* **2010**, *81*, 174118.

[30]    Y. Ivry, D. P. Chu, J. F. Scott, C. Durkan, *Phys. Rev. Lett.* **2010**, *104*, 207602.

[31]    Y. Ivry, D. P. Chu, C. Durkan, *Nanotechnology* **2010**, *21*, 065702.

[32]    R. Xu, S. Liu, I. Grinberg, J. Karthik, A. R. Damodaran, A. M. Rappe, L. W. Martin, *Nat. Mater.* **2015**, *14*, 79.

[33]    I. A. Khan, X. Marti, C. Serrao, R. Ramesh, S. Salahuddin, *Nano Lett.* **2015**.

[34]  P. Paruch, J. Guyonnet, *Comptes Rendus Phys.* **2013**, *1*, 1.

[35]  Y. Ivry, C. Durkan, D. Chu, J. F. Scott, *Adv. Funct. Mater.* **2014**, *24*, 5567.

[36]  B. Tummers, DataTheif III version **2006**.

[37]  G.-T. Park, J.-J. Choi, J. Ryu, H. Fan, H.-E. Kim, *Appl. Phys. Lett.* **2002**, *80*, 4606.

[38]  Y. Ivry, D. Chu, J. F. Scott, E. K. H. Salje, C. Durkan, *Nano Lett.* **2011**, *11*, 4619.

[39]  D. Wu, I. Vrejoiu, M. Alexe, A. Gruverman, *Appl. Phys. Lett.* **2010**, *96*, 112903.

[40]  C. Durkan, D. P. Chu, P. Migliorato, M. E. Welland, *Appl. Phys. Lett.* **2000**, *76*, 366.

[41]  Y. Ivry, D. Chu, C. Durkan, *Appl. Phys. Lett.* **2009**, *94*, 162903.

[42]  V. Nagarajan, A. Roytburd, A. Stanishevsky, S. Prasertchoung, T. Zhao, L. Chen, J. Melngailis, O. Auciello, R. Ramesh, *Nat. Mater.* **2003**, *2*, 43.

[43]  G. Le Rhun, I. Vrejoiu, M. Alexe, *Appl. Phys. Lett.* **2007**, *90*, 012908.

[44]  V. Anbusathaiah, D. Kan, F. C. Kartawidjaja, R. Mahjoub, M. a. Arredondo, S. Wicks, I. Takeuchi, J. Wang, V. Nagarajan, *Adv. Mater.* **2009**, *21*, 3497.

[45]  G. Arlt, *Ferroelectrics* **1987**, *76*, 451.

[46]  N. Balke, I. Bdikin, S. V. Kalinin, A. L. Kholkin, *J. Am. Ceram. Soc.* **2009**, *92*, 1629.

[47]  C. Durkan, J. A. Garcia-Melendrez, L. Ding, *J. electromechanical sensors* **2015**, *44*, 2230.

[48]  F. Jona, G. Shirane, *Ferroelectric Crystals*; Dover Publications, 1993.

[48]  C. Harnagea, M. Alexe & D. Hesse, *IEEE Transactions on Ultrasonics, Ferroelectrics & Frequency Control*; **53**, 2309, 2006.

[49]  B. J. Rodriguez, C. Callahan, S. V. Kalinin & R. Proksch, *Nanotechnology*, **18**, 475504 2007.

[50]  F. Jona, G. Shirane, *Ferroelectric Crystals*; Dover Publications, 1993.

[51]  I. Horcas, R. Fernández, J. M. Gómez-Rodríguez, J. Colchero, J. Gómez-Herrero, A. M. Baro, *Rev. Sci. Instrum.* **2007**, *78*, 013705.

**Supplementary Information to:** *Towards resolving Landauer's Paradox through direct observation of the multiscale ferroelastic-ferroelectric interplay*

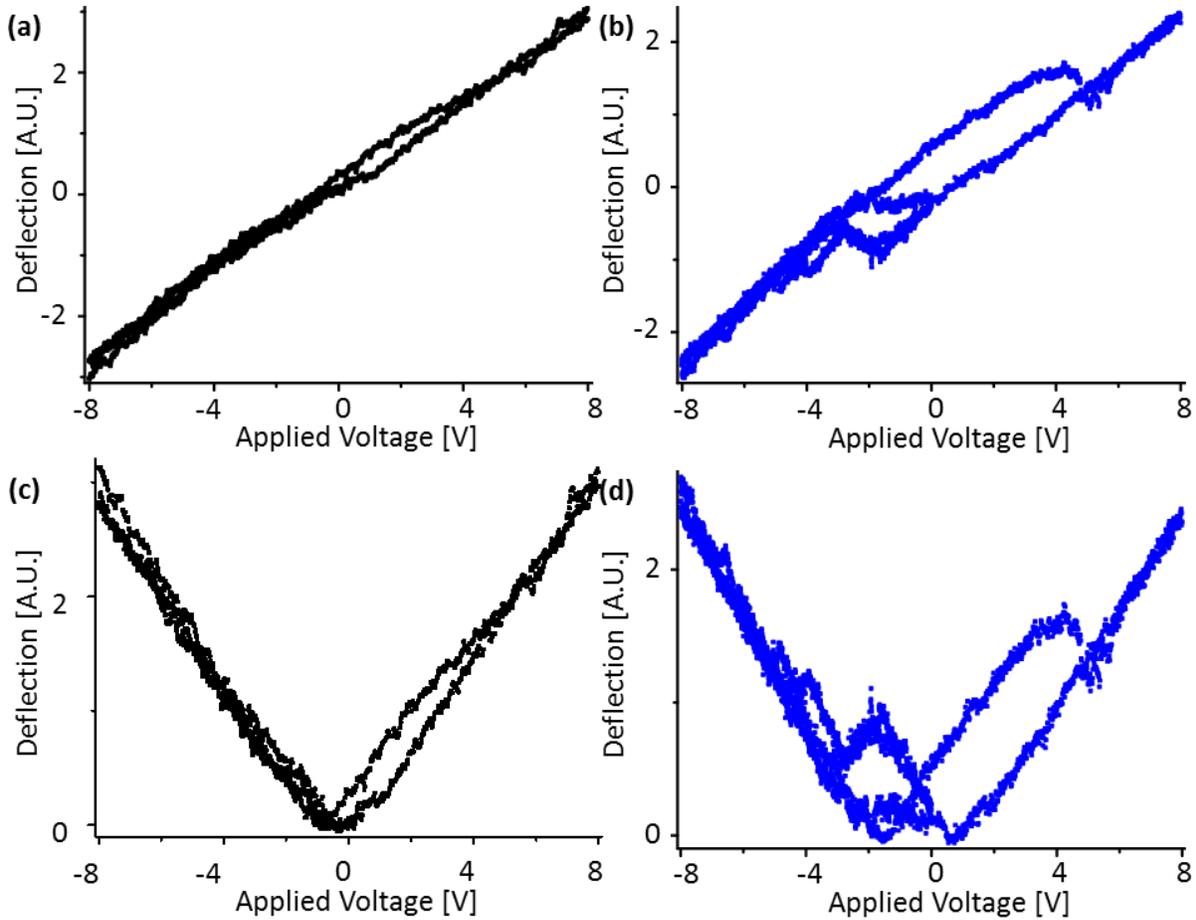

**Figure SI1 LHLM of ferroelectric and of ferroelastic domain switching**. In-phase PFM LHLM of (a) purely ferroelectric domain switching and of (b) ferroelastic domain switching, and the 'butterfly' (amplitude PFM) LHLM of the amplitude signal of (c) purely ferroelectric domain switching and of (d) ferroelastic domain switching. (a) and (c) are the LHLM done at point 'A' in Fig. 1 and (b) and (d) are taken from point 'B' in Fig. 1.